\documentclass{PoS}
\bibliography{my-bib-database}
\usepackage{color}
\usepackage{wrapfig,lipsum,booktabs}
\def\1{\c{c}}
\def\2{\c{C}}
\def\3{\.{I}}
\def\4{\"{a}}
\def\5{{\i}}
\def\6{$\beta$}
\def\7{\"{o}}
\def\8{\"{O}}
\def\9{\c{s}}
\def\0{\c{S}}
\def\*{\"{u}}
\def\?{\"{U}}
\def\;{\u{g}}
\def\:{\u{G}}

\title{A New Gamma-Ray Source in the Vicinity of the Galactic Supernova Remnant G306.3$-$0.9}

\ShortTitle{A New Gamma-Ray Source in the Vicinity of the Galactic Supernova Remnant G306.3$-$0.9}

\author{\speaker{T\*l\*n Ergin}
        TUBITAK Space Technologies Research Institute, Ankara, Turkey\\
        E-mail: \email{tulun.ergin@tubitak.gov.tr}}

\author{Satoru Katsuda,
        Saitama University, Saitama, Japan}

\author{Aytap Sezer,
        Avrasya University, Trabzon, Turkey}

\author{Ryo Yamazaki,
        Aoyama Gakuin University, Fuchinobe, Japan}
 
 \author{Miroslav Filipovic,
       Western Sydney University, Sydney, Australia}
          
\author{Hidetoshi Sano,
       Nagoya University, Nagoya, Japan}

\author{Yasuo Fukui,
       Nagoya University, Nagoya, Japan}

\author{Shuta Tanaka,
       Konan University, Kobe, Japan}

\abstract{A new extended gamma-ray source, which was named as Source A, in the southwest of Galactic supernova remnant (SNR) G306.3$-$0.9 was detected with a significance of  $\sim$13$\sigma$ at the location of R.A. (J2000) = 13$^{\rm{h}}$ 17$^{\rm{m}}$ 52$^{\rm{s}\!\!}$.80, Decl. (J2000) = $-$63$^{\circ}$ 55$'$ 48$''\!\!$.00 using about 9 years of {\it Fermi}-LAT data. In order to investigate this unidentified gamma-ray source in multi-wavelengths, we performed {\it Swift} observations of Source A. In this presentation we summarize the published gamma-ray results, report about the recent ToO {\it Swift} observations of Source A, and show our preliminary results of the gamma-ray analysis that we conducted using the new X-ray data.}

\FullConference{7th Fermi Symposium 2017\\
		15-20 October 2017\\
		Garmisch-Partenkirchen, Germany}
		
\begin{document}

\section{Introduction}
\vspace{-0.5cm}
In \cite{Se17}, we reported about the detection of an extended gamma-ray source, Source A, located $\sim$0$^{\circ}\!\!$.6 south-west of the supernova remnant (SNR) G306.3$-$0.9. Assuming Source A as a point-like gamma-ray source,we detected it with a significance of $\sim$9.7$\sigma$ (TS\footnote{Test Statistics (TS) values indicate that the null hypothesis (maximum likelihood value for a model without an additional source) is incorrect. The square-root of TS gives the detection significance of a source.}$\sim$94). Its best-fitted location was found to be R.A. (J2000) = 13$^{\rm{h}}$ 17$^{\rm{m}}$ 52$^{\rm{s}\!\!}$.80, Decl. (J2000) = $-$63$^{\circ}$ 55$'$ 48$''\!\!$.00. If the extended gamma-ray emission was fit to a disk-like extension model, the extension radius was measured to be 0$^{\circ}\!\!$.73 $\pm$ 0$^{\circ}\!\!$.07. As an extended source the total significance was found to be $\sim$13$\sigma$ and assuming a power-law (PL) spectrum, we obtained $\Gamma$ = 2.1 and the energy flux was found to be (2.07 $\pm$ 0.2) $\times$ 10$^{-5}$ MeV cm$^{-2}$ s$^{-1}$. 

In this paper we summarize the results of our {\it Swift} ToO observations on Source A in Section 2. Following the ToO observations, we extended the gamma-ray analysis by using 3 months of more {\it Fermi}-LAT data than what we used in our previous analysis \cite{Se17}. In Section 3, we explain the gamma-ray analysis and give its preliminary results. In Section 4, we present the conclusions and give an outlook. 

\vspace{-0.5cm}
\section{{\it Swift} ToO Observations \& Results}
\vspace{-0.5cm}
To unravel the nature of the extended unidentified Fermi gamma-ray source, Source A, found near the SNR G306.3$-$0.9 \cite{Se17}, two {\it Swift} ToOs (IDs: 00010121001, 00010151001, 00010151002) were successfully completed in May and June 2017. We had 5.6 ks effective exposure for the May observations and 5 ks for June observations. Four new X-ray sources were discovered, which we named as SrcA/Src1, SrcB/Src2, SrcC/Src3, and SrcD/Src4 in our initial analyses. The June ToO was centered at SrcB, because this was found to be the brightest X-ray source. Except SrcC, all {\it Swift} XRT sources are within the 5$\sigma$ contour level of Source A.

\begin{figure}
\centering
\includegraphics[width=0.4\textwidth]{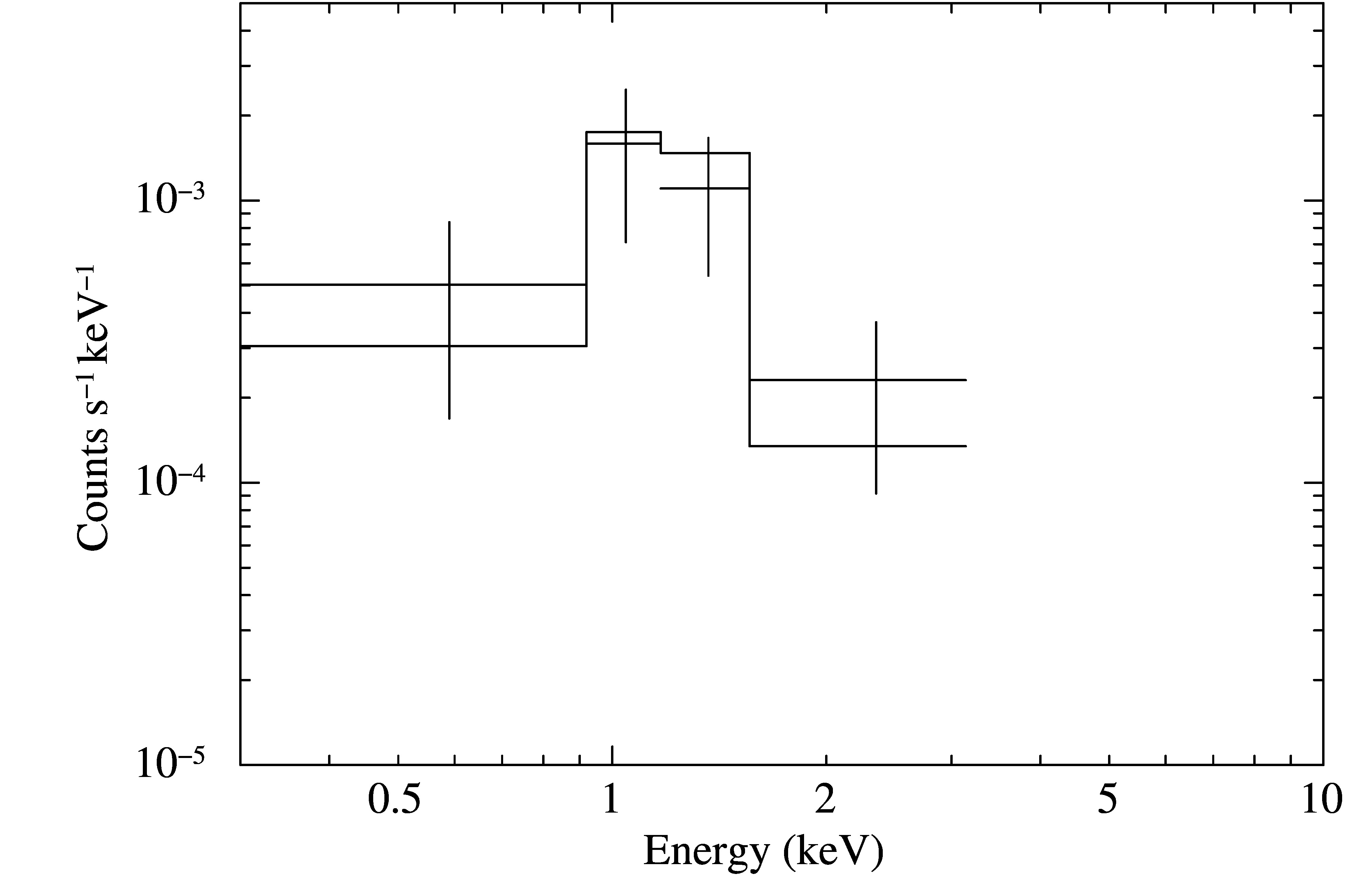}
\includegraphics[width=0.4\textwidth]{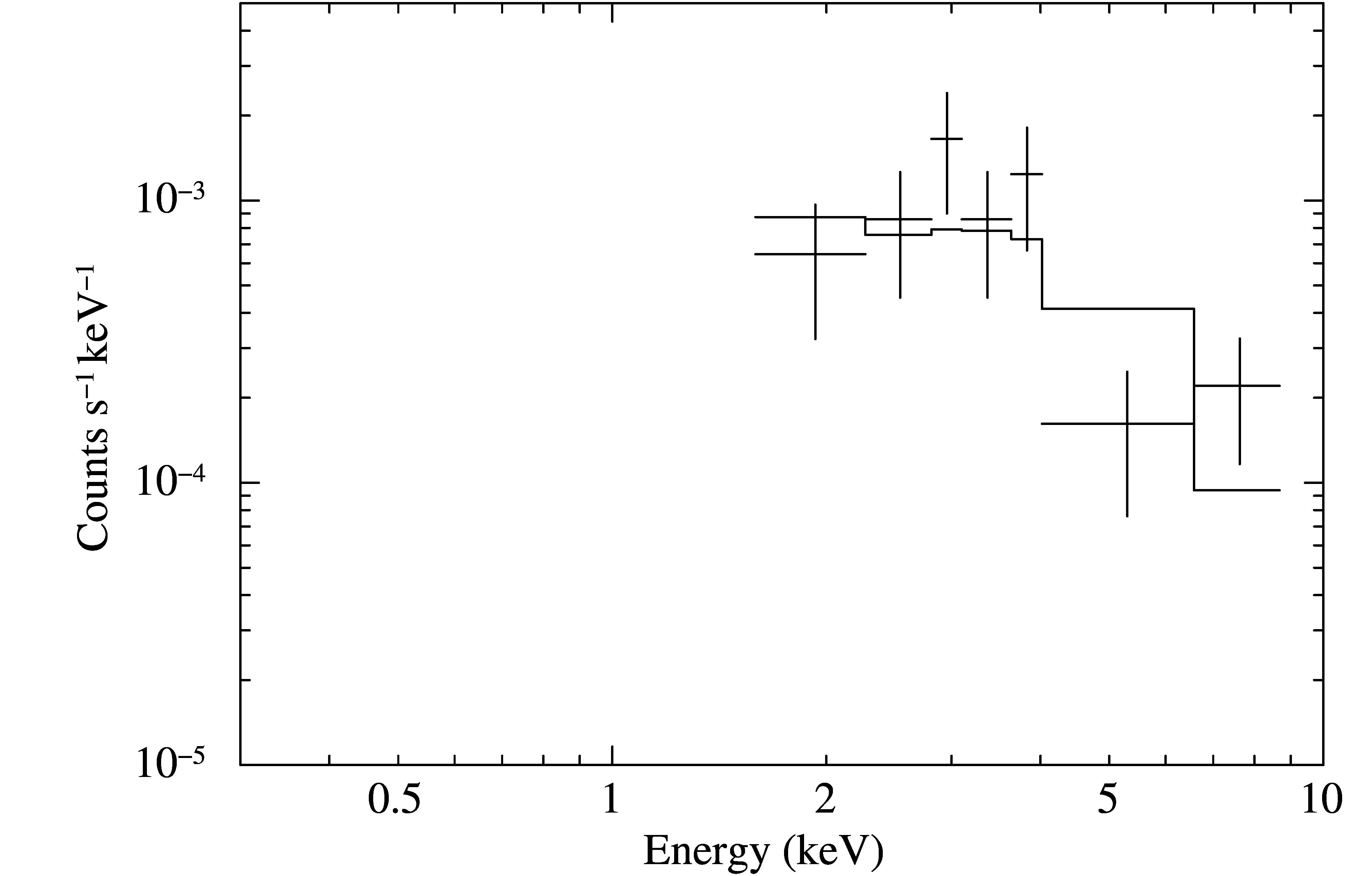}
\includegraphics[width=0.4\textwidth]{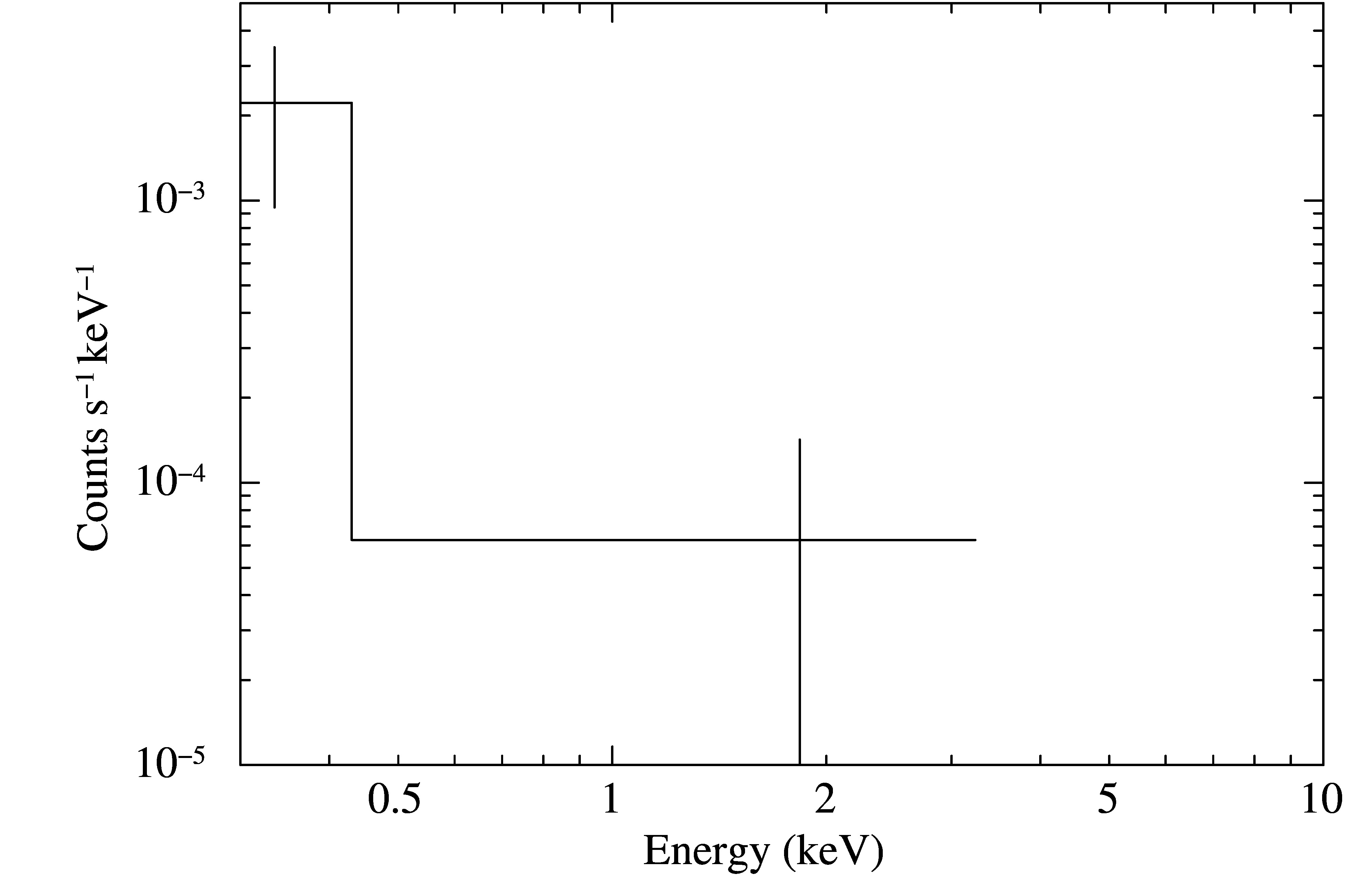}
\includegraphics[width=0.4\textwidth]{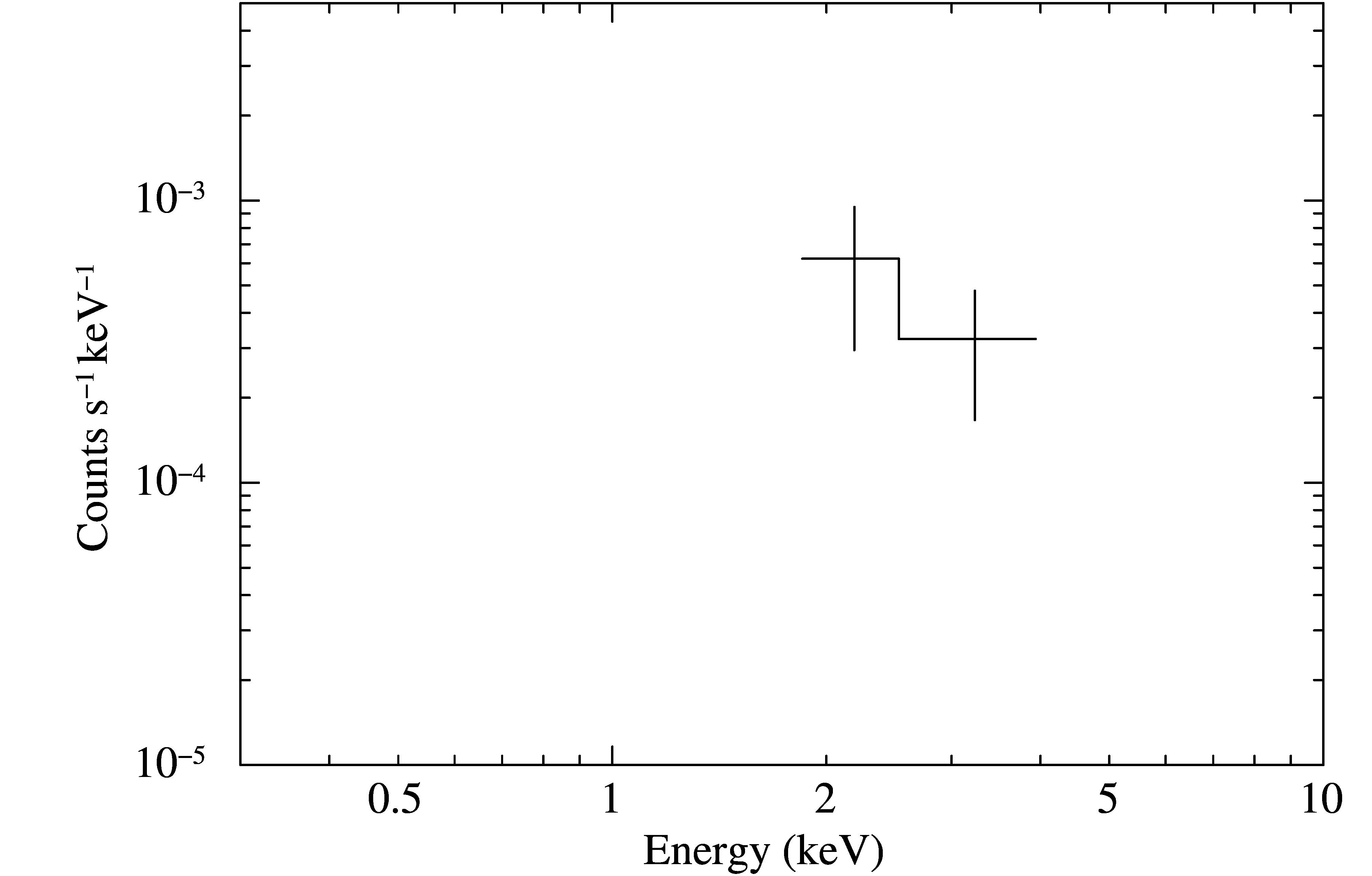}
\vspace{-0.2cm}
\caption{ \footnotesize{{\it Swift} XRT spectra of SrcA (upper-left panel), SrcB (upper-right panel), SrcC (lower-left panel), and SrcD (lower-right panel). The photon indices are allowed to vary freely and they are found to be 8.6 for SrcA, 1.7 for SrcB, 5.9 for SrcC, 2.8 for SrcD.}}
\label{figure_1}
\vspace{-0.5cm}
\end{figure}

\begin{wraptable}{r}{0.55\textwidth}
\vspace{-0.7cm}
  \caption{ X-ray point sources observed by {\it Swift} XRT in two ToO observations.}
   \vspace{0.2cm}
\begin{tabular}{@{}lccccc@{}}
  \hline
\hline
Name  & RA (deg) & Dec. (deg) & Exposure (ks)     \\
\hline
SrcA/Src1 &  199.331 & -63.882   &10.6  \\
SrcB/Src2 &  199.742 & -63.954   & 10.6  \\
SrcC/Src3 &  199.579  & -63.781  &10.6  \\
SrcD/Src4 &  198.964  & -63.905  & 10.6 \\
  \hline
\end{tabular}
\label{table_1}
\vspace{-0.3cm}
\end{wraptable}
The locations of these X-ray sources are given in Table \ref{table_1}. The {\it Swift} XRT sources are about 10$'$ away from each other. So, there is no physical connection between them, because the separation between them is too big for any distance that is $>$0.1 kpc. All {\it Swift} XRT sources, are close to or within the 5$\sigma$ contour level of Source A. Here are the details of the initial {\it Swift} analysis results and the multi-wavelength aspects of each of these four X-ray sources: 
\vspace{-0.3cm}
\begin{itemize}
\item {\bf SrcA (Src1):} This is the closest {\it Swift} XRT source to Source A. SrcA has an optical counterpart classified as a star in the Guide Star Catalog 2.3\footnote {http://gsss.stsci.edu/Catalogs/GSC/GSC2/GSC2.htm}, which is 4$''$ (S7KT124582) away from SrcA. Another close optical source (S7KT125052) is 6$''$ away from SrcA. SrcA might be a binary system, because a separation of $\sim$6$''$ amounts to $\sim$6000 AU at a distance of 1 kpc. SrcA has no radio counterparts. 
\vspace{-0.3cm}
\item {\bf SrcB (Src2):} Flux of SrcB changed by a factor of 4 within 3 weeks implying that this source is a variable. However, there are no observed optical or radio counterparts for this source.\vspace{-0.3cm}
\item {\bf SrcC (Src3):} This source has a very soft spectrum and since its position is outside the 5$\sigma$ gamma-ray contours of  Source A, we assume that this source is not directly related to Source A.
\vspace{-0.3cm}
\item {\bf SrcD (Src4):} This source was found to exist in the {\it Swift} XRT data, while searching for radio counterparts for SrcA, SrcB, and SrcC in the Sydney University Molonglo Sky Survey (SUMSS) data. We found a SUMSS radio counterpart for SrcD at 843 MHz that also overlaps with PMN J1315-6354 (R.A. (J2000) = 13$^{\rm{h}}$ 15$^{\rm{m}}$ 52$^{\rm{s}\!\!}$.9, Decl. (J2000) = $-$63$^{\circ}$ 54$'$ 40$''$) from the Parkes-MIT-NRAO (PMN) 4.85GHz Surveys catalog \cite{Wr94}. 
\end{itemize}
\vspace{-0.2cm}
X-ray spectra for all {\it Swift} XRT sources are shown in Figure \ref{figure_1}. For SrcA, SrcB, and SrcD we assumed the absorbing Hydrogen column density to be 1.7 $\times$ 10$^{22}$ cm$^{-2}$, which is the same as that for G306.3$-$0.9, to calculate the absorption corrected flux in the 0.5-10 keV energy range. Hydrogen column density for SrcC was assumed to be zero to adequately fit its very soft X-ray spectrum. The absorption corrected flux value for SrcA, SrcB, SrcC, and SrcD is calculated to be $\sim$3.2 $\times$ 10$^{-11}$, $\sim$4.5 $\times$ 10$^{-13}$, $\sim$2.5 $\times$ 10$^{-15}$, $\sim$3.3 $\times$ 10$^{-13}$ ergs cm$^{-2}$ s$^{-1}$, respectively. Since SrcB is the brightest X-ray source among all four detected {\it Swift} XRT sources, we calculated the luminosity of SrcB to be $\sim$6 $\times$ 10 $^{31}$ erg s$^{-1}$ at a distance of 1 kpc.

\begin{wrapfigure}{r}{0.47\textwidth} 
\vspace{-27pt}
\begin{center}
\includegraphics[width=0.47\textwidth]{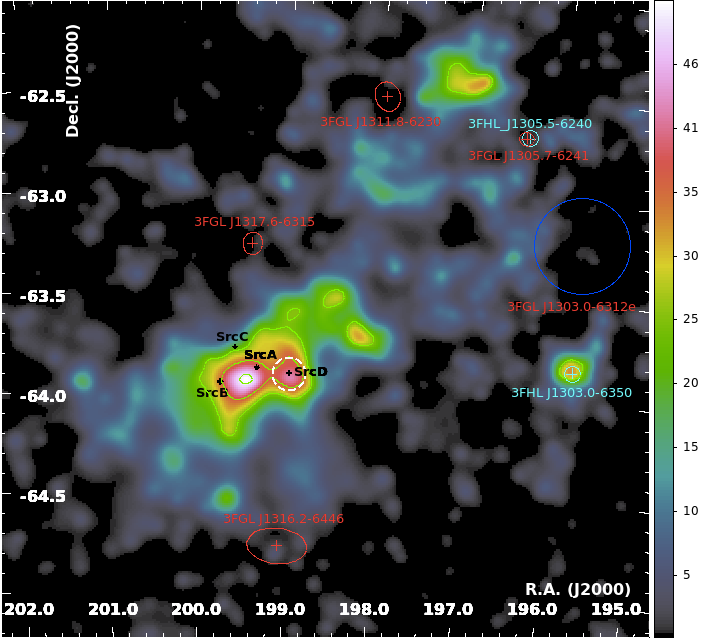}
\vspace{-26pt}
\caption{\footnotesize{The gamma-ray TS map of Source A, which is not included in the background model.}}
\label{figure_1a}
\end{center}
\vspace{-27pt}
\end{wrapfigure}

We checked the 3rd Fermi-LAT source catalog (3FGL) \cite{Ac15} and the 3rd catalog of hard Fermi-LAT sources (3FHL) \cite{Ac17} to find possible counterparts for Source A and the associated X-ray sources. In Figure \ref{figure_1a} {\it Swift} X-ray sources are shown with black markers. Fermi-LAT sources from the 3rd Fermi-LAT source catalog are shown in red markers. Sources from the 3FHL catalog are shown in cyan markers. The extended 3FGL source (3FGL J1303.0-6312e) was also reported in the 3FHL catalog and its extension is shown as a blue circle. The green contours are for the gamma-ray TS values (25, 36, 49). We could not find any counterparts for Source A in 3FGL and 3FHL catalogs. The white dashed circle shows the radio source PMN J1315-6354.

\vspace{-0.5cm}
\section{Analysis \& Results of Gamma-ray Data}
\begin{figure}
\vspace{-0.3cm}
\centering
\includegraphics[width=0.45\textwidth]{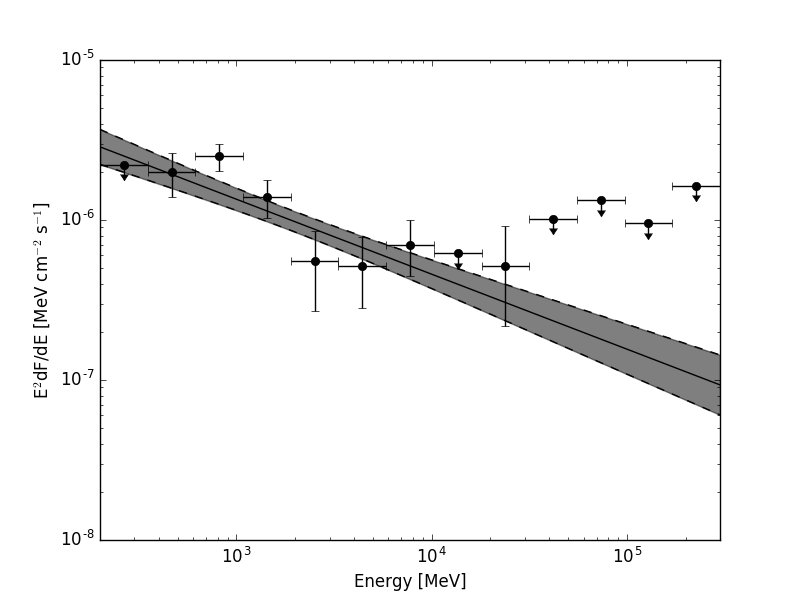}
\includegraphics[width=0.45\textwidth]{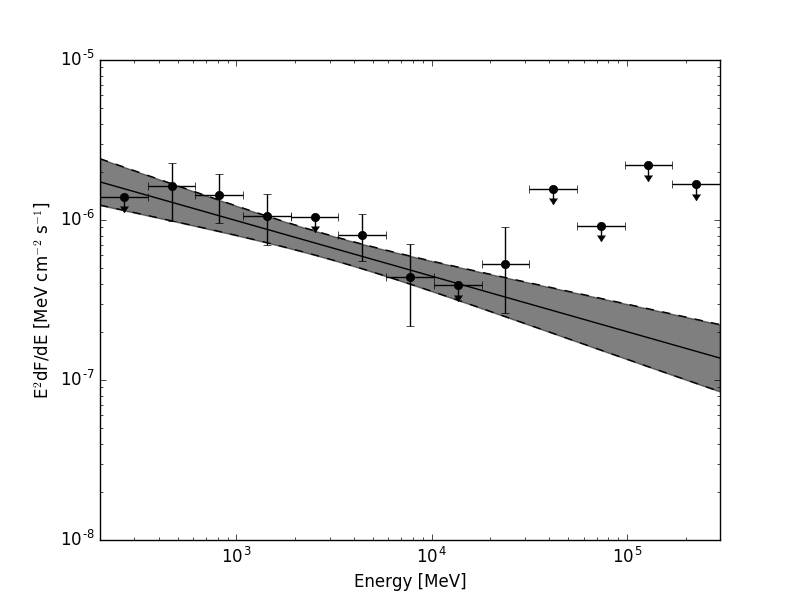}
\vspace{-0.2cm}
\caption{ \footnotesize{The SED of SrcB (left panel) and SrcD (right panel) assuming a PL-type spectrum in the energy range of 0.2$-$300 GeV for both sources. The central solid black line and grey band represent the best-fitted PL model and its statistical errors. }}
\label{figure_3}
\vspace{-0.3cm}
\end{figure}

\begin{figure}
\vspace{-0.3cm}
\centering
\includegraphics[width=0.45\textwidth]{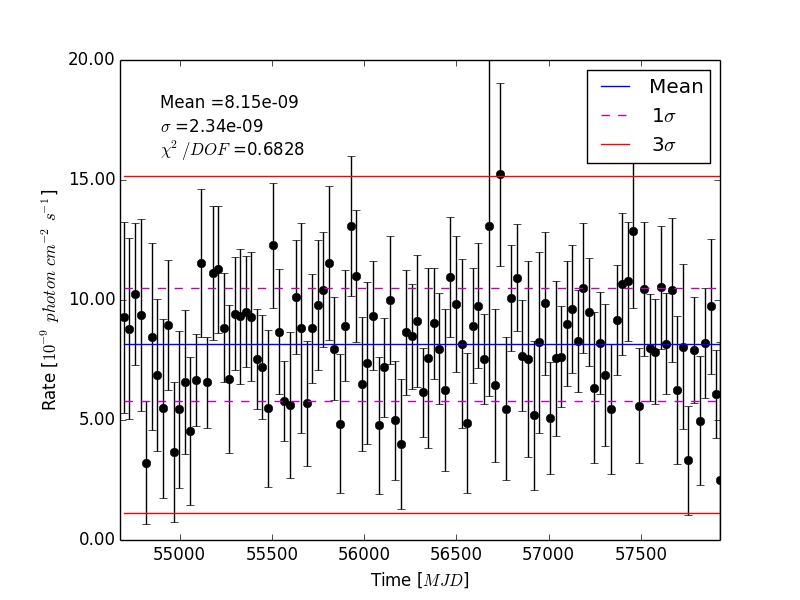}
\includegraphics[width=0.45\textwidth]{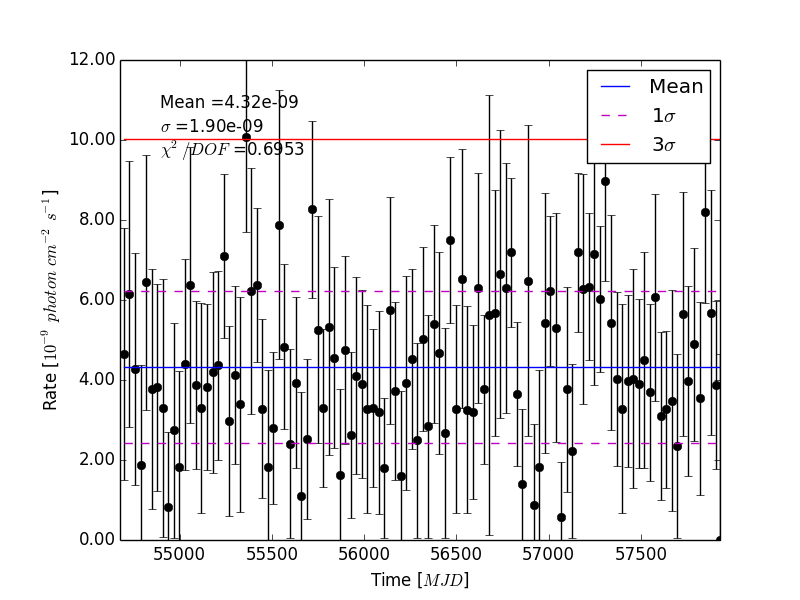}
\vspace{-0.2cm}
\caption{ \footnotesize{The 1-month-binned gamma-ray light curve of SrcB (left panel) and SrcD (right panel) assuming a PL-type spectrum in the energy range of 0.2$-$300 GeV for both sources. The blue line shows the mean value. The dashed magenta and solid red line represents the 1$\sigma$ and 3$\sigma$ significance levels, respectively.}}
\label{figure_4}
\vspace{-0.3cm}
\end{figure}

\vspace{-0.5cm}
After discovering the X-ray sources within Source A, we re-analyzed the gamma-ray data. We used data between 2008-08-04 and 2017-06-30. We analyzed events within the energy range of 200 MeV - 300 GeV using the Fermi analysis toolkit \texttt{fermipy}\footnote{http://fermipy.readthedocs.io/en/latest/index.html}. We selected Fermi-LAT Pass 8 `Source' class and front+back type events, which come from zenith angles smaller than 90$^{\circ}$ and  from within a circular region of interest (ROI) with a radius of 20$^{\circ}$ centered at the best-fit position of Source A. The maximum likelihood fitting method was employed on the spatially and spectrally binned data and used the instrument function P8R2$_{-}$SOURCE$_{-}\!\!$V6. The gamma-ray background model contains Galactic diffuse sources ({\it gll$_{-}$iem$_{-}\!$v6.fits}) and isotropic sources ({\it iso$_{-}$P8R2$_{-}$SOURCE$_{-}\!\!$V6$_{-}\!$v06.txt}). It also includes all point-like and extended sources from the 3rd Fermi-LAT Source Catalog located within a 15$^{\circ}\times$15$^{\circ}$ region centered at the ROI center. Freed normalization parameters of sources that are within 3$^{\circ}$ of ROI center. Freed all parameters of the diffuse Galactic emission and the isotropic component. All sources with TS $>$ 10 are set free and all sources with TS $<$ 10 are fixed. 
The analysis region shown by the 10$^{\circ}\times$10$^{\circ}$ TS map covers a very large area in the sky and Source A seems to show some sub-structures, but the X-ray sources concentrated around the best-fitted location of Source A. In order to clarify how much each X-ray source is contributing to Source A's total gamma-ray emission, we added each of the X-ray sources one by one as a point-like source into the gamma-ray background model. Then we checked for the significances and produced a TS map for every version of the gamma-ray background model. Since SrcC is located out of the 5$\sigma$ contours of Source A, we assumed that it is not a part of Source A. Thus, we excluded SrcC from the gamma-ray analysis. 
\begin{table}
 \begin{minipage}{150mm}
  \begin{center}
  \caption{The maximum likelihood analysis results for different combinations of {\it Swift} XRT sources and G306.3$-$0.9 that are included in the gamma-ray background model.}
   \vspace{-0.2cm}
\begin{tabular}{@{}lccccc@{}}
  \hline
\hline
Included Source Names 			    &  TS (SrcA) & TS (SrcB) & TS (SrcD) & TS (G306.3$-$0.9) \\
\hline
G306.3$-$0.9              			    &  -           	 &  -                 & -                 & 55.48\\
SrcA \& G306.3$-$0.9 			    &  128.92       &  -                 & -                 & 14.42\\
SrcB \& G306.3$-$0.9 			    &  -                 & 139.10        & -                 &   6.14\\
SrcD \& G306.3$-$0.9 			    &  -  	          & -                  & 111.89       & 27.50 \\
SrcA \& SrcD \& G306.3$-$0.9 		    &   89.29        & -                  &   24.26       &  6.08\\
SrcA \& SrcB \& G306.3$-$0.9 		    &   42.57        &  46.86         & -                 & 8.15\\
SrcB \& SrcD \& G306.3$-$0.9 		    &  -                 &  74.09         &   50.59       & 9.13\\
SrcA \& SrcB \& SrcD \& G306.3$-$0.9 &    2.52         &  90.81         &   43.27       & 3.73\\
  \hline
\end{tabular}
\label{table_2}
\end{center}
\end{minipage}
\vspace{-0.3cm}
\end{table}
\begin{figure}
\centering
\includegraphics[width=0.40\textwidth]{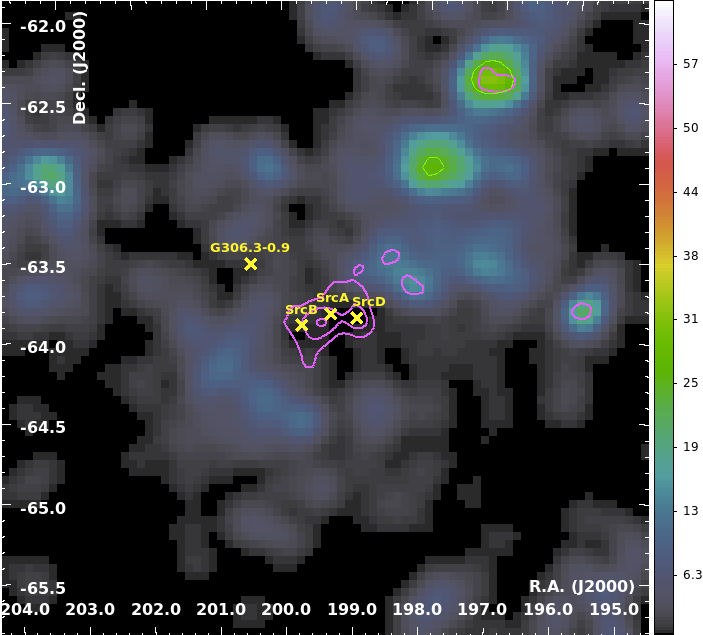}
\includegraphics[width=0.40\textwidth]{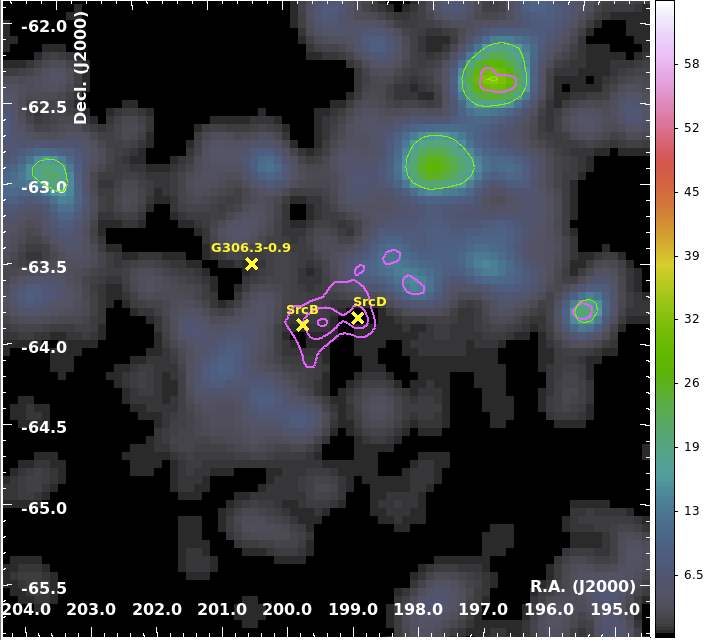}
\vspace{-0.2cm}
\caption{ \footnotesize{The TS map for two different background gamma-ray models. Left Panel: Including SrcA, SrcB, SrcD, and G306.3$-$0.9 in the background model. Right Panel: Including SrcB, SrcD, and G306.3$-$0.9 in the background model. On both panels, the magenta significance contours of 5, 6, and 7$\sigma$ are taken from Figure \ref{figure_1a}. All sources added to the background model are shown with yellow crosses. Green contours represent the 5$\sigma$ significance level obtained after the background model is fit to the data.}}
\label{figure_2}
\vspace{-0.5cm}
\end{figure}
We found out that the included source combination of 'SrcB \& SrcD \& G306.3$-$0.9' cleans all excess gamma-ray emission from the nearby region of the best-fitted position of Source A. Although including 'SrcA \& SrcB \& SrcD \& G306.3$-$0.9' source combination also gives comparable results to the 'SrcB \& SrcD \& G306.3$-$0.9' combination, the TS value of SrcA comes out as 2.52 in the former analysis. The TS map with SrcA, SrcB, SrcD, and G306.3$-$0.9 included in the gamma-ray background model is shown in Figure \ref{figure_2} left panel and the same figure right panel shows the TS map for only including SrcB, SrcD, and G306.3$-$0.9 in the gamma-ray background model. 
A PL spectral fit to the spectra of SrcB and SrcD gives  a spectral index of  2.47 $\pm$ 0.12 and 2.35 $\pm$ 0.14, respectively, and a total energy flux of (5.9 $\pm$ 1.1) $\times$ 10$^{-6}$ MeV cm$^{-2}$ s$^{-1}$ and (4.6 $\pm$ 1.0) $\times$ 10$^{-6}$ MeV cm$^{-2}$ s$^{-1}$, respectively. The spectra of SrcB and Src D are shown in Figure \ref{figure_3}.
To see the long term variability in the light curve of SrcB and SrcD, we apply Fermi-LAT aperture photometry taking data from the circular region of 1$^{\circ}$ around the best-fitting position of SrcB and SrcD. For each source we applied the barycenter correction to the data. We also applied event weighing to keep only events with a probability of $>$ 10\% of being from SrcB/SrcD. Higher probability values decreases the number of events abruptly. The 1-month-binned light curves of SrcB and SrcD assuming a PL-type spectrum in the energy range of 0.2 - 300 GeV are shown on the left and right panels of Figure \ref{figure_4}.

\vspace{-0.5cm}
\section{Conclusions \& Outlook}
\vspace{-0.5cm}
We analyzed the GeV gamma-ray data including the recently detected {\it Swift} XRT sources and the SNR G306.3$-$0.9 in the gamma-ray background as point-source source templates. Comparing the analysis results for the case where 'SrcA \& SrcB \& SrcD \& G306.3$-$0.9' source combination was part of the background model with the ones for 'SrcB \& SrcD \& G306.3$-$0.9' source combination inside the background model, we found comparable results. However, by looking at the excess in the TA maps, 'SrcB \& SrcD \& G306.3$-$0.9' source combination is favored. Possible source-type scenarios for SrcA, SrcB, SrcD are listed as follows: 
\vspace{-0.3cm}
\begin{itemize}
\item SrcA has an optical counterpart, which is a star. It could be a gamma-ray binary, and if so, its variability has to be observed in various wavelengths. However, it is not detected in gamma rays (TS=2.52).
\vspace{-0.3cm}
\item SrcB is the brightest X-ray source among all {\it Swift} XRT sources. It has no optical and radio counterpart, but was detected in gamma rays with a significance of $\sim$9$\sigma$. This source is possibly a quasar with a very weak radio emission, but probably the SUMSS sensitivity limit is insufficient for the detection of SrcB. Optical and radio observations are needed to confirm this estimate.
\vspace{-0.3cm}
\item SrcD has a radio counterpart found in SUMSS. It was detected in gamma rays with a significance of $\sim$7$\sigma$. It may be a blazar candidate, but more observations are needed to show the variability in different wave-bands. 
\end{itemize}
\vspace{-0.3cm}
As a next step, we plan for multi-waveband observations (radio, optical, and X-rays) on SrcB and SrcD. In addition, we will re-analyze the gamma-ray data  to do more variability checks and investigate the energy-dependent source morphology.


\end{document}